\documentstyle[10pt,psfig,mathtools,amssymb ,twocolumn,xcolor,flushend,cite]{IEEEtran}

\newcommand{\bi}{\begin{itemize}}
\newcommand{\ei}{\end{itemize}}

\newcommand{\be}{\begin{enumerate}}
\newcommand{\ee}{\end{enumerate}}
\newcommand{\bd}{\begin{description}}
\newcommand{\ed}{\end{description}}
\newcommand{\bc}{\begin{center}}
\newcommand{\ec}{\end{center}}
\newcommand{\bt}{\begin{tabbing}}
\newcommand{\et}{\end{tabbing}}
\newcommand{\bfig}{\begin{figure}}
\newcommand{\efig}{\end{figure}}
\newcommand{\beq}{\begin{equation}}
\newcommand{\beqarr}{\begin{eqnarray}}
\newcommand{\beqarrn}{\begin{eqnarray*}}
\newcommand{\eeq}{\end{equation}}
\newcommand{\eeqarr}{\end{eqnarray}}
\newcommand{\eeqarrn}{\end{eqnarray*}}
\newcommand{\bflr}{\begin{flushright}\vspace{-0.2in}}
\newcommand{\eflr}{\end{flushright}}
\newcommand{\bsub}{\begin{subequations}}
\newcommand{\esub}{\end{subequations}}
\newcommand{\barr}{\begin{array}}
\newcommand{\earr}{\end{array}}
\newcommand{\nn}{\nonumber}

\def\binom#1#2{\left( \! \! \barr{c} #1 \\ #2 \earr \! \! \right)}

\def\undb#1{\mbox{\bf{#1}}}

\begin{document}
\title{\huge{Optimal One- and Two-Sided Multi-Level ASK for Noncoherent SIMO Systems Over Correlated Rician Fading}
}

\author{Badri Ramanjaneya Reddy, Soumya P. Dash,~\IEEEmembership{Senior Member,~IEEE},\\ and George C. Alexandropoulos,~\IEEEmembership{Senior Member,~IEEE}
\thanks{B. R. Reddy and G. C. Alexandropoulos are with the Department of Informatics and Telecommunications, National and Kapodistrian University of Athens, 16122 Athens, Greece. G. C. Alexandropoulos is also with the Department of Electrical and Computer Engineering, University of Illinois Chicago, Chicago, IL 60601, USA (e-mails: \{badrirreddy, alexandg\}@di.uoa.gr).}
\thanks{S. P. Dash is with the School of Electrical Sciences, Indian Institute of Technology Bhubaneswar, Argul, Khordha, Odisha, India 752050 (e-mail: soumyapdashiitbbs@gmail.com).}
\vspace{-1.0cm}
% \thanks{Copyright (c) 2015 IEEE. Personal use of this material is permitted. However, permission to use this material for any other purposes must be obtained from the IEEE by sending a request to pubs-permissions@ieee.org.}
}
\maketitle
%%%%%%%%%%%%%%%%%%%%%%%%%%%%%%%%%%%%%%%%%%%%%%%%%%%%%%%%%%%%%%%%%%%%%%%%%%%%%%%%%%%%%%%%%%%%%
\begin{abstract}
This paper analyzes the performance of a single-input multiple-output (SIMO) wireless communication system employing one- and two-sided amplitude shift keying (ASK) modulation schemes for data transmission and operating under correlated Rician fading channels. The receiver deploys an optimal noncoherent maximum likelihood detector, which exploits statistical knowledge of the channel state information for signal decoding. An optimal receiver structure is derived, from which series-form and closed-form expressions for the union bound on the symbol error probability (SEP) are obtained for general and massive SIMO systems, respectively. Furthermore, an optimization framework to derive the optimal one- and two-sided ASK modulation schemes is proposed, which focuses on minimizing SEP performance under an average transmit energy constraint. The conducted numerical investigations for various system parameters demonstrate that the proposed noncoherent SIMO system with the designed optimal ASK modulation schemes achieves superior error performance compared to traditional equispaced ASK modulation. It is also shown that, when the proposed system employs traditional two-sided ASK modulation, superior error performance from the case of using one-sided ASK is obtained.
\end{abstract}
%%%%%%%%%%%%%%%%%%%%%%%%%%%%%%%%%%%%%%%%%%%%%%%%%%%%%%%%%%%%%%
%%%%%%%%%%%%%%%%%%%%%%%%%%%%%%%%%%%%%%%%%%%%%%%%%%%%%%%%%%%%%%%%%%%%%%%%%%%%%%%%%%%%%%%%%%%%%%
%%%%%%%%%%%%%%%%%%%%%%%%%%%%%%%%%%%%%%%%%%%%%%%%%%%%%%%%%%%%%%%%%%%%%%%%%%%%%%%%%%%%%%%%%%%%%%
%%%%%%%%%%%%%%%%%%%%%%%%%%%%%%%%%%%%%%%%%%%%%%%%%%%%%%%%%%%%%%%%%%%%%%%%%%%%%%%%%%%%%%%%%%%%%%
%%%%%%%%%%%%%%%%%%%%%%%%%%%%%%%%%%%%%%%%%%%%%%%%%%%%%%%%%%%%%%%%%%%%%%%%%%%%%%%%%%%%%%%%%%%%%%
%\vspace{-0.8cm}
\section{Introduction} 
\IEEEPARstart{O}{ne} of the core use cases of the fifth Generation (5G) of wireless networks is Ultra-Reliable and Low-Latency Communications (URLLC) \cite{GhMaBa19, AlNoAb23, AlZiBa21}. This use case is specifically intended to meet the stringent demands of applications that require both low latency and high reliability, making it essential for mission-critical operations. To achieve these requirements, spatial diversity combined with coherent detection at the receiver has been extensively studied \cite{OsLaDu21, ZeLvLi20, RePaDe20}. Coherent communications rely on the availability of Instantaneous Channel State Information (CSI) at the receiver, which requires dedicated estimation techniques that typically use training/pilot symbols. However, accurate CSI estimation necessitates long pilot overhead~\cite{XuIsRa19}. 

In URLLC applications, data packets are usually small, which implies that the length of the training signals is limited. As a result, the deployment of training sequences may lead to significant overhead, while their short length may severely degrade the quality of channel estimation. Furthermore, adding a long pilot sequence to those small packets results in longer processing times, which increases latency. To address these issues, noncoherent communications are typically adopted, which do not rely on instantaneous CSI. Instead, a noncoherent communication system uses statistical CSI at the receiver to decode the received data. This approach has been applied in URLLC applications to reduce both complexity and latency~\cite{GaZhCh19}.

Over the years, noncoherent communications have also gained increasing interest due to their advantages of low complexity, reduced power consumption, and minimized signaling requirements, which eliminates the need for pilot symbols in the channel estimation process. As a result, this form of communications has been applied across various domains and applications, with extensive studies being devoted to its performance analysis. Notable applications include ambient backscatter communications \cite{GuLuHo20, DeDh21}, power line communications \cite{KuDa21}, and Reconfigurable Intelligent Surfaces (RISs) \cite{Ca23,ris1,ris2}. For instance, a machine learning framework to design a noncoherent transceiver over a massive Single-Input and Multiple-Output (SIMO) multipath channel was proposed in \cite{ZhLaHu20}. In \cite{SuZhDo23}, the authors developed a noncoherent SIMO framework with modulation on conjugate-reciprocal zeros for short packet transmissions, proposed a noncoherent Viterbi-like detector, and studied the system's error performance. It was shown that the use of optimal constellations leads to improved system error performance. In~\cite{ChMaGo16}, a noncoherent massive SIMO system was studied where the receiver uses statistical CSI to decode the received data, with optimal constellations further enhancing system error performance. In \cite{XiXuNg20}, the authors proposed a joint multi-user constellation in an energy detection-based noncoherent massive multiple-input multiple-output system over Rayleigh fading channels. Furthermore, in \cite{MaMu14}, a noncoherent SIMO system was analyzed using the Amplitude Shift Keying (ASK) modulation scheme for data transmission over Rayleigh fading. The study showed that better error performance is achieved when the transmitter uses ASK symbols in geometric progression rather than traditional arithmetic progression. This analysis was extended to correlated Rayleigh fading channels in \cite{MaMuDa16}. In \cite{DuNgBe23}, the authors designed low-complexity multi-level constellations based on Kullback-Leibler divergence for noncoherent SIMO systems, which enhanced error performance. Finally, in \cite{LiDoCh21}, an optimal constellation to improve symbol detection reliability for short data packet transmission over noncoherent massive SIMO Rayleigh fading channels was designed. 

As can be verified from the latter studies, the state-of-the-art analyses on noncoherent multi-antenna communication systems have largely overlooked the presence of a Line-of-Sight (LoS) component in signal transmissions. However, in 5G and beyond, dense wireless network deployments play a crucial role in enhancing signal coverage and service quality \cite{HeYuAn21, PaWuZh24}. In such environments, the reduced distance between the communicating devices minimizes the impact of obstructions, such as tall buildings and rough terrain, making the LoS paths more dominant than the non-LoS ones. Consequently, the likelihood of a LoS component in the transmitted signal increases, and a widely accepted channel model for such fading conditions is the Rician distribution~\cite{JaLoDi18}. 

Noncoherent communications with ASK signals over uncorrelated Rician fading channels was only recently studied in~\cite{ReDaGh24, ReDaLi24}, where an optimization framework for ASK modulation schemes was proposed. Improved error performance compared to conventional ASK schemes was demonstrated. Optimal one- and two-sided ASK modulation schemes were presented for RIS-aided noncoherent systems in~\cite{MuReDa24, MiDaAl24}, showcasing that the proposed schemes achieve superior error performance compared to their traditional counterparts. However, in practical wireless transmissions, spatial correlation arises due to limited antenna spacing, making it a critical factor in channel modeling~\cite{FoNgMa19, PeAlDa11, MaAlNg12, AK2015}. This correlation is influenced by both physical constraints and propagation characteristics (e.g., low angular spread~\cite{MAD2023}). Motivated by these considerations and the research gap, this paper investigates a noncoherent SIMO system employing ASK modulation over correlated Rician fading channels, incorporating both spatial correlation among diversity branches and the presence of a LoS component. The major contributions of the paper are summarized as follows:
 \begin{itemize}
     \item An optimal noncoherent Maximum Likelihood (ML) detector is presented for which a tailored receiver structure is designed.
     \item Utilizing the derived receiver structure, we present a series-form expression for a union bound on the Symbol Error Probability (SEP) performance. 
     \item We devise an optimization framework to obtain optimal ASK modulation schemes by minimizing the SEP under an average transmit energy constraint.
     \item Extensive numerical investigations that also verify our analytical expressions through comparisons with equivalent simulations are presented to demonstrate the superior error performance achieved with the use of the proposed optimal ASK modulation schemes for data transmission, as compared to traditional equispaced ASK constellations.  
 \end{itemize}

The rest of the paper is organized as follows. Section II outlines the system model and introduces the proposed receiver structure. Section III presents a novel analytical expression in the form of a series for the union bound on the SEP, while Section IV discusses our optimization framework to determine the optimal ASK modulation schemes that minimize the SEP under an average transmit energy constraint. In Section V, numerical evaluations are presented to validate our SEP analysis and illustrate the structure of the optimal modulation schemes. Section VI concludes the paper listing our final remarks.

%%%%%%%%%%%%%%%%%%%%%%%%%%%%%%%%%%%%%%%%%%%%%%%%%%%%%%%%%%%%%%%%%%%%%%%%%%%%%%%%%%%%%%%%%%%%%%
%%%%%%%%%%%%%%%%%%%%%%%%%%%%%%%%%%%%%%%%%%%%%%%%%%%%%%%%%%%%%%%%%%%%%%%%%%%%%%%%%%%%%%%%%%%%%%
%%%%%%%%%%%%%%%%%%%%%%%%%%%%%%%%%%%%%%%%%%%%%%%%%%%%%%%%%%%%%%%%%%%%%%%%%%%%%%%%%%%%%%%%%%%%%%
\section{System and Channel Models}
We consider a wireless communication system in which a single-antenna transmitter sends data symbols on a symbol-by-symbol basis over correlated Rician fading channels. The receiver receives the transmitted data symbols with $N$ antennas, and symbol-by-symbol detection is performed. The received complex symbol at the $\ell$-th diversity branch (with $\ell = 1, \ldots, N$) in baseband representation is given by
\beq
r_{\ell} \triangleq h_{\ell}s + n_{\ell},
\label{eq1}
\eeq
where $s$ is the information bearing symbol and $n_{\ell}$ is the complex-valued additive noise at the $\ell$-th diversity branch. The $N \times 1$ complex-valued additive noise vector $\undb{n} \triangleq \left [n_1, \ldots, n_N  \right ]^{\rm T}$ follows a zero-mean complex Gaussian distribution, i.e., $\undb{n} \sim {\mathcal{CN}}\left(\undb{0}_{N \times 1}, \sigma_n^2 \undb{I}_N \right)$, where $\undb{0}_{N \times 1}$ is the $N \times 1$ zeros vector, $\undb{I}_N$ is the $N \times N$ identity matrix, and $\sigma_n^2$ is the variance of each element of $\undb{n}$. In addition, $h_{\ell}$ is the complex channel fading gain at the $\ell$-th diversity branch. The $N \times 1$ complex-valued channel fading gain vector $\undb{h} \triangleq \left [h_1, \ldots, h_N\right ]^{\rm T}$ follows a non-zero mean complex Gaussian distribution, i.e., $\undb{h} \sim {\mathcal{CN}} \left( \pmb{\mu}, \undb{K}_h\right)$, where $\pmb{\mu} \triangleq \left [\mu_1, \ldots, \mu_N \right ]^{\rm T}$ and $\undb{K}_h \triangleq \mathbb{E}\left [ \left(\undb{h} - \pmb{\mu}\right)\left(\undb{h}-\pmb{\mu}\right)^{\rm H}\right ]$ are the $N \times 1$ mean vector and the $N \times N$ covariance matrix of $\undb{h}$, respectively. Here, $\left[ \cdot\right]^{\rm T}$ denotes matrix transposition, $\left[ \cdot\right]^{\rm H}$ denotes the Hermitian (conjugate transpose) operator, and $\mathbb{E}\left [ \cdot\right ]$ is the expectation operator. 

We have considered the following three cases for the channel covariance matrix $\undb{K}_h$ in this paper:
\begin{enumerate}
    \item {\em Independent and identically distributed (i.i.d.)
channels}:
    In this case, $\undb{K}_h$ is a diagonal matrix with identical elements along the diagonal. The element located at each $i$-th row and $j$-th column (with $i, j = 1, \ldots, N$) is given by the expression:
    \beqarr
        \left(\undb{K}_h \right)_{ij} = \begin{cases}
            \sigma_h^2, & \text{if}\,\, i = j\\  
 0, & \text{if}\,\, i \neq j 
        \end{cases}.
        \label{eq2}
    \eeqarr
    \item {\em Uniformly correlated channels~\cite{AlSaLa09}}:
    In this case, $\undb{K}_h$ is an uniformly correlated matrix and its element in each $i$-th row and $j$-th column is given by:
    \bsub
    \beqarr
        \left(\undb{K}_h \right)_{ij} = \begin{cases}
            \sigma_h^2, & \text{if}\,\, i = j\\  
 \sigma_h^2\epsilon, & \text{if}\,\, i \neq j
        \end{cases},
        \label{eq3a}
    \eeqarr
        where it should hold that:
        \beq
        -\frac{1}{N-1} < \epsilon < 1,\,\,\epsilon \neq 0.
        \label{eq3b}
        \eeq
    \esub
    \item {\em Exponentially correlated channels~\cite{AlSaLa09, AlMaSa10, AlPe18}}:  In this case, $\undb{K}_h$ is an exponentially correlated matrix and its element in the $i$-th row and $j$-th column is given by:
    \bsub
    \beqarr
        \left(\undb{K}_h \right)_{ij} = \begin{cases}
            \sigma_h^2 & \text{if}\,\, i = j\\  
 \sigma_h^2\epsilon^{|i-j|} & \text{if}\,\, i \neq j 
        \end{cases},
        \label{eq4a}
    \eeqarr
        where 
        \beq
        -1 < \epsilon < 1,\,\,\epsilon \neq 0.
        \label{eq4b}
        \eeq
    \esub
\end{enumerate}

The eigendecomposition of $\undb{K}_h$ is given by
%\beq
$\undb{K}_h = \sigma_h^2\undb{U}\pmb{\Lambda}\undb{U}^{\rm H}$
%\label{eq5}
%\eeq
where $\sigma_h^2$ represents the average variance of the elements of $\undb{h}$, i.e., $\sigma_h^2 = \text{tr}\left(\undb{K}_h\right)/N$, with $\text{tr}(\cdot)$ denoting the trace of a matrix. The matrix $\undb{U}$ is a unitary matrix with columns corresponding to the orthonormal eigenvectors of $\undb{K}_h$, and $\pmb{\Lambda}$ is a diagonal matrix with diagonal elements $\lambda_1, \ldots, \lambda_N$. These elements represent the eigenvalues of the normalized channel covariance matrix $\left(1/\sigma_h^2 \right)\undb{K}_h$. It is also noted that:
\begin{enumerate}
    \item In the case of {\em i.i.d. channels}, all eigenvalues of the normalized covariance matrix are equal to one, i.e., $\lambda_1 = \ldots = \lambda_N = 1$.
    \item In the case of {\em uniformly correlated channels}, the eigenvalues of the normalized covariance matrix are $1-\epsilon$ with a multiplicity of $N - 1$, and $1 + \left( N-1 \right)\epsilon$ with a multiplicity of 1.
    \item In the case of {\em exponentially correlated channels}, all the eigenvalues of the normalized covariance matrix are distinct.
\end{enumerate}

In this paper, we assume that the information-bearing symbol $s$ in~\eqref{eq1} is chosen from a multi-level ASK modulation scheme. More specifically, we consider two types of multi-level ASK modulation schemes, each with $M$ levels. The $m$-th symbol in these modulation schemes is given by:  
\begin{enumerate}
\bsub
\item {\em One-sided ASK}:
\beqarr
s_m &=&  \sqrt{E_m} ,\,\,\, m = 1,  \ldots, M, \nn \\
\sqrt{E_m} &<& \sqrt{E_{m+1}},\,\, m = 1, \ldots M-1.
\label{eq5a}
\eeqarr
\item {\em Two-sided ASK}:
\beq
s_m = \begin{cases} 
 -\sqrt{E_{\frac{M}{2} - m +1}}, & \text{if}\,\, 1\leq m \leq \frac{M}{2}\\  
 \sqrt{E_{ m - \frac{M}{2}}}, & \text{if}\,\, \frac{M}{2} < m \leq M 
 \end{cases}, 
 \label{eq5b}
\eeq
\esub
\end{enumerate}
where $E_m$ represents the energy of each $m$-th symbol in the modulation scheme. Assuming that the symbols are equiprobable, the average energy of the modulation scheme, which is denoted by $E_{av}$, is given as follows:
\beq
E_{av} \triangleq \frac{1}{M} \sum_{m = 1}^M s_m^2.
\label{eq6}
\eeq
Considering that the receiver has perfect statistical knowledge of the channel, the received signal vector $\undb{r}$ can be processed by multiplying it with the matrix $\undb{U}\in{\mathbb{C}}^{ N\times N}$, resulting in:
\beq
\widetilde{\undb{r}} \triangleq \widetilde{\undb{h}}s + \widetilde{\undb{n}},
\label{eq7}
\eeq
where $\widetilde{\undb{r}} \triangleq \undb{U}^{\rm H}\undb{r}$, $\widetilde{\undb{h}} \triangleq \undb{U}^{\rm H}\undb{h}$, and $\widetilde{\undb{n}}\triangleq\undb{U}^{\rm H}\undb{n}$. Here, $\widetilde{\undb{h}} \sim {\mathcal{CN}} \left(\undb{U}^{\rm H}\pmb{\mu}, \sigma_h^2\pmb{\Lambda} \right)$ and $\widetilde{\undb{n}}\sim {\mathcal{CN}} \left(\undb{0}_{N \times 1}, \sigma_n^2 \undb{I}_N \right)$. Thus, the distribution of $\widetilde{\undb{r}}$ conditioned on $s$ is given by:
\beq
\widetilde{\undb{r}}|_s \sim {\mathcal{CN}}\left(\widetilde{\pmb{\mu}}s, |s|^2\sigma_h^2\pmb{\Lambda} + \sigma_n^2\undb{I}_N \right),
\label{eq8}
\eeq
where $\widetilde{\pmb{\mu}} \triangleq \undb{U}^{\rm H} \pmb{\mu}$. 

Using the above, the average SNR per symbol per branch, denoted by $\Gamma_{av}$, can be expressed as follows: 
\beq
\Gamma_{av} \triangleq \frac{1}{M} \sum_{m=1}^M \frac{s_m^2 \sigma_h^2}{\sigma_n^2} =  \frac{1}{M} \sum_{m=1}^M \Gamma_m= \frac{E_{av} \sigma_h^2}{\sigma_n^2},
\label{eq9}
\eeq
where $\Gamma_m$ is the average SNR per branch of each $m$-th symbol, i.e., $
\Gamma_m = \frac{s_m^2 \sigma_h^2}{\sigma_n^2}.$ The multi-antenna receiver can then decode the received signal using a noncoherent ML detector, leveraging the statistical knowledge of the channel. Thus, its decision rule is derived as:
\beq
\hat{s} \triangleq \arg \max_{s} \,\, \ln f\left(\widetilde{\undb{r}}|_s\right) \, ,
\label{eq10}
\eeq
where $f\left(\widetilde{\undb{r}}|_s\right)$ denotes the probability density function (p.d.f.) of the modified received signal vector $\widetilde{\undb{r}}$ conditioned on the data symbol $s$. By substituting the p.d.f. of $\widetilde{\undb{r}}|_s$ in (\ref{eq10}), the simplified decision rule can be expressed as follows:
\beqarr
\hat{s} &=& \arg \min_{s} \,\, \left \{ \left(\widetilde{\undb{r}}-s\widetilde{\pmb{\mu}}\right)^{\rm H}\left(\left |s \right|^2\sigma_h^2\pmb{\Lambda} + \sigma_n^2\undb{I}_N\right)^{-1}\left(\widetilde{\undb{r}}-s\widetilde{\pmb{\mu}}\right) \right. \nn \\
&&\qquad \qquad +\left. \ln \left( \left| \left|s\right|^2\sigma_h^2\pmb{\Lambda} + \sigma_n^2\undb{I}_N \right|\right) \right\} \nn \\
&=& \arg \min_{s} \sum_{\ell =1}^N \frac{\left| \widetilde{r}_{\ell} - s \widetilde{\mu}_{\ell}\right|^2}{\left | s\right|^2 \sigma_h^2\lambda_{\ell}+\sigma_n^2} + \ln \left(\left | \left | s\right |^2\sigma_h^2\pmb{\Lambda}+\sigma_n^2\undb{I}_N\right |\right).\nn \\
\label{eq11}
\eeqarr
The latter expression will be next used to analyze the performance of the considered noncoherent multi-antenna receiver under correlated Rician fading conditions.
%%%%%%%%%%%%%%%%%%%%%%%%%%%%%%%%%%%%%%%%%%%%%%%%%%%%%%%%%%%%%%%%%%%%%%%%%%%%%%%%%%%%%%%%%%%%%%
%%%%%%%%%%%%%%%%%%%%%%%%%%%%%%%%%%%%%%%%%%%%%%%%%%%%%%%%%%%%%%%%%%%%%%%%%%%%%%%%%%%%%%%%%%%%%%
%%%%%%%%%%%%%%%%%%%%%%%%%%%%%%%%%%%%%%%%%%%%%%%%%%%%%%%%%%%%%%%%%%%%%%%%%%%%%%%%%%%%%%%%%%%%%%
\section{Symbol Error Probability Analysis}
The union bound on the SEP performance of the aforedescribed receiver can be obtained using the Pairwise Error Probability (PEP), which is mathematically expressed as:
\beq
P_{e, UB} \leq \frac{1}{M} \sum_{i=1}^M \sum_{\substack{j=1, \\ j\neq i}}^M P_{i \rightarrow j}.
\label{eq12}
\eeq

Let us assume, without loss of generality, that the $i$-th symbol of the modulation scheme, $s_i$, has been transmitted. Using the receiver structure described in (\ref{eq11}), the probability of erroneously detecting it as $s_j$ can be expressed as:
\beqarr
&&\!\!\!\!\!\! \! \! \! \! \!
P_{i \rightarrow j} \nn \\
&=& \Pr \left ( \sum_{\ell =1}^N \frac{\left| \widetilde{r}_{i, \ell} - s_i \widetilde{\mu}_{\ell}\right|^2}{\left | s_i\right|^2 \sigma_h^2\lambda_{\ell}+\sigma_n^2} + \ln \left(\left | \left | s_i\right |^2\sigma_h^2\pmb{\Lambda}+\sigma_n^2\undb{I}_N\right |\right) \right. \nn \\
&&\quad>\!\! \left. \sum_{\ell =1}^N \frac{\left| \widetilde{r}_{i, \ell} - s_j \widetilde{\mu}_{\ell}\right|^2}{\left | s_j\right|^2 \sigma_h^2\lambda_{\ell}+\sigma_n^2} + \ln \left(\left | \left | s_j\right |^2\sigma_h^2\pmb{\Lambda}+\sigma_n^2\undb{I}_N\right |\right)\right ). \nn \\
\label{eq13}
\eeqarr
By using the definition $d_{i,\ell} \triangleq \left | s_i\right|^2 \sigma_h^2\lambda_{\ell}+\sigma_n^2$, the latter expression can be re-written as:
\beqarr
&&\!\!\!\!\!\! \! \!
P_{i \rightarrow j} \nn \\
&&\!\!\!\!\!\!= \Pr \left ( \sum_{\ell =1}^N \frac{\left| \widetilde{r}_{i, \ell} - s_j \widetilde{\mu}_{\ell}\right|^2}{d_{j, \ell}}-\frac{\left| \widetilde{r}_{i, \ell} - s_i \widetilde{\mu}_{\ell}\right|^2}{d_{i, \ell}} < \ln \frac{\prod\limits_{\ell = 1}^N d_{i, \ell}}{\prod\limits_{\ell = 1}^N d_{j, \ell}}\right ),\nn \\
\label{eq14}
\eeqarr
which can be further expanded as:
\beqarr
&&\!\!\!\!\!\!\!\!\!\!P_{i \rightarrow j} \nn \\
&=& \Pr \left ( \sum_{\ell =1}^N \left [\frac{d_{i, \ell} -d_{j, \ell}}{d_{i, \ell}d_{j, \ell}} \left | \widetilde{r}_{i, \ell} - s_i \widetilde{\mu}_{\ell} + \frac{d_{i, \ell} \left(s_i-s_j \right)}{d_{i, \ell} -d_{j, \ell}}\widetilde{\mu}_{\ell}\right |^2 \right. \right.\nn \\
&&\qquad\qquad\quad\,\,\, \left.-\frac{\left(s_i - s_j \right)^2\left|\widetilde{\mu}_{\ell}\right|^2}{d_{i, \ell} -d_{j, \ell}} \right ]< \left.\ln \frac{\prod\limits_{\ell = 1}^N d_{i, \ell}}{\prod\limits_{\ell = 1}^N d_{j, \ell}}\right )\!.
\label{eq15}
\eeqarr
Using the distribution of $\widetilde{\undb{r}}|_s$ given in (\ref{eq8}) yields $\widetilde{r}_{i, \ell} = \sqrt{d_{i, \ell}}z_{\ell} + s_i \widetilde{\mu}_{\ell}$ $\forall i,\ell$, where $z_{\ell}$ is the standard normal random variable, i.e., $z_{\ell} \sim {\mathcal{CN}}\left(0, 1 \right)$. Therefore, (\ref{eq15}) results in:
\beqarr
&&\!\!\!\!\!\!\!\!\!\!\!\!\!\!\!\!\!\!\!\!P_{i \rightarrow j} \nn \\
&=& \Pr \left ( \sum_{\ell =1}^N \frac{d_{i, \ell} -d_{j, \ell}}{d_{j, \ell}} \left | z_{\ell}  + \frac{\sqrt{d_{i, \ell}} \left(s_i-s_j \right)}{d_{i, \ell} -d_{j, \ell}}\widetilde{\mu}_{\ell}\right |^2 \right.\nn \\
&&\qquad\quad< \left.\ln \frac{\prod\limits_{\ell = 1}^N d_{i, \ell}}{\prod\limits_{\ell = 1}^N d_{j, \ell}} + \sum_{\ell =1}^N \frac{\left(s_i - s_j \right)^2\left|\widetilde{\mu}_{\ell}\right|^2}{d_{i, \ell} -d_{j, \ell}}\right )\!. 
\label{eq16}
\eeqarr

We now use the following definitions in \eqref{eq16}:
\beqarr
\chi_{ij}^2 &\triangleq& \sum_{\ell =1}^N \frac{d_{i, \ell} -d_{j, \ell}}{d_{j, \ell}} \left | z_{\ell}  + \frac{\sqrt{d_{i, \ell}} \left(s_i-s_j \right)}{d_{i, \ell} -d_{j, \ell}}\widetilde{\mu}_{\ell}\right |^2,\nn \\
\alpha_{ij} &\triangleq& \ln \frac{\prod\limits_{\ell = 1}^N d_{i, \ell}}{\prod\limits_{\ell = 1}^N d_{j, \ell}} + \sum_{\ell =1}^N \frac{\left(s_i - s_j \right)^2\left|\widetilde{\mu}_{\ell}\right|^2}{d_{i, \ell} -d_{j, \ell}}.
\label{eq17}
\eeqarr
Clearly, the former random variable follows the noncentral chi-square distribution. Furthermore, considering $L$ distinct eigenvalues for $\undb{K}_h$ and that each $p$-th eigenvalue has multiplicity of $q_p$, the moment generating function (m.g.f.) of $\chi_{ij}^2$ can be expressed as follows:
\beq
G_{ij} \left( \nu\right) \triangleq \prod\limits_{p=1}^L \frac{\exp \left \{ \frac{\nu \left(\frac{d_{i, p} -d_{j, p}}{d_{j, p}}\right) \sum \limits_{q=1}^{q_p}\left |\frac{\sqrt{d_{i, p}} \left(s_i-s_j \right)}{d_{i, p} -d_{j, p}}\widetilde{\mu}_{q,p} \right |^2}{1-\nu\left(\frac{d_{i, p} -d_{j, p}}{d_{j, p}}\right)}\right\}}{\left(1-\nu\left(\frac{d_{i, p} -d_{j, p}}{d_{j, p}}\right) \right)^{q_p}}.
\label{eq18}
\eeq
Utilizing the SNR expressions given in (\ref{eq9}), the latter m.g.f. and $\alpha_{ij}$ can be written as:
\beqarr
G_{ij} \left( \nu\right) &=& \prod\limits_{p=1}^L \frac{\exp \left \{ \frac{\nu \beta_{ij,p} \gamma_{ij, p} \sum\limits_{q=1}^{q_p}k_{q,p}}{1-\nu\beta_{ij,p}}\right\}}{\left(1-\nu\beta_{ij,p} \right)^{q_p}}, \nn \\
\alpha_{ij} &=& \ln \frac{\prod\limits_{p = 1}^L \left( \Gamma_i\lambda_p + 1 \right)^{q_p}}{\prod\limits_{p = 1}^L \left( \Gamma_j\lambda_p + 1 \right)^{q_p}} \nn \\
&&+ \sum\limits_{p = 1}^L \frac{\left(\sqrt{\Gamma_i\lambda_p}\phi_i - \sqrt{\Gamma_j\lambda_p}\phi_j \right)^2}{\Gamma_i\lambda_p - \Gamma_j\lambda_p}\sum_{q=1}^{q_p} k_{q,p},
\label{eq19}
\eeqarr
where we have used the definitions:
\beqarr
\beta_{ij,p} &\triangleq& \frac{\Gamma_i\lambda_p+1}{\Gamma_j\lambda_p+1} -1,\nn \\
\gamma_{ij, p} &\triangleq& \left(\Gamma_{i}\lambda_p + 1 \right)\left(\frac{\sqrt{\Gamma_i\lambda_p}\phi_i - \sqrt{\Gamma_j\lambda_p}\phi_j}{\Gamma_i\lambda_p - \Gamma_j\lambda_p} \right)^2,
\label{eq20}
\eeqarr
with $k_{q,p} = \frac{\left | \widetilde{\mu}_{q,p}\right |^2}{\sigma_h^2\lambda_p}$ $\forall q = 1, \ldots, q_p$. We also define the average Rician factor as $K_{av} \triangleq \sum\limits_{p=1}^L\sum\limits_{q=1}^{q_p} \frac{k_{q,p}}{N}$. 

By applying~\cite[Lemma 1]{RaMo19}, a tight approximation for the cumulative distribution function (c.d.f.) of $\chi_{ij}^2$ can be obtained from its m.g.f. as follows:
\beqarr
F_{\chi_{ij}^2}\left( x \right) \approx \sum_{u = 0}^{\xi-1} \frac{\left(\xi-1 \right)^u}{x^u u!} \frac{d^u}{d\nu^u} G_{ij} \left( \nu\right)\Bigg |_{\nu = \frac{1-\xi}{x}}, 
\label{eq21}
\eeqarr
where $\xi$ is a large integer. Utilizing \cite[Lemma 3.2b.1]{PrMa92}, it holds that:
\beq
\frac{d^u}{d\nu^u} G_{ij} \left( \nu\right) = G_{ij} \left( \nu\right)R_u\left( \nu\right),
\label{eq22}
\eeq
where 
\begin{align}\label{eq23}
R_u\left( \nu\right) &\triangleq \sum_{v=0}^{u-1}\binom{u-1}{v}g^{(u-1-v)}\left( \nu\right) R_v\left(\nu \right),\\
g^{(n)}\left( \nu\right)\!\! &\triangleq \!\!\sum_{p=1}^L \frac{n!\left(\left(1+n\right)\gamma_{ij,p}\sum\limits_{q=1}^{q_p}k_{q,p} \!\!+ q_p\left(1-\beta_{ij,p}\nu\right)\right)}{\beta_{ij,p}^{-n-1}\left(1-\beta_{ij,p}\nu \right)^{n+2}}.\nonumber
\end{align}
Owing to the structure of the constellation given in (\ref{eq5a}) and (\ref{eq5b}) and the previously derived c.d.f. expression in (\ref{eq21}), the PEP in (\ref{eq16}) can be further simplified for the cases: \textit{i}) $d_{i, \ell} > d_{j, \ell}$; \textit{ii}) $d_{i, \ell} < d_{j, \ell}$; and \textit{iii}) $d_{i, \ell} = d_{j, \ell}$ for $s_i =- s_j$. For the former two cases, an analytical expression for the PEP performance can be obtained as follows:
\beqarr
\!\!\!\!\!\!\!\!\!\!P_{i \rightarrow j} \!&=&\! \Pr \left( \chi_{ij}^2 < \alpha_{ij} \right) \!=\! F_{\chi_{ij}^2}\left( \alpha_{ij} \right)\!, \,\,\,\,\,\,\,\,\,\,\,\,\text{for}\,\,  d_{i, \ell} \!<\! d_{j, \ell}\nn \\
\!\!\!\!\!\!\!\!\!\!P_{i \rightarrow j} \!&=&\! \Pr \left( \chi_{ij}^2 > \alpha_{ij} \right) \!=\! 1- F_{\chi_{ij}^2}\left( \alpha_{ij} \right)\!,\,\, \text{for}\,\,  d_{i, \ell}\! > \!d_{j, \ell}.
\label{eq24}
\eeqarr
For the case of $d_{i, \ell} = d_{j, \ell}$ for $s_i =- s_j$, the $P_{i \rightarrow j}$ can be expressed using (\ref{eq14}) as
\beqarr
P_{i \rightarrow j} &=& \Pr \left ( \sum_{\ell =1}^N \frac{1}{d_{i, \ell}} \left(\left| \widetilde{r}_{i, \ell} + s_i \widetilde{\mu}_{\ell}\right|^2-\left| \widetilde{r}_{i, \ell} - s_i \widetilde{\mu}_{\ell}\right|^2\right) < 0\right ) \nn \\
&=& \Pr \left ( \sum_{\ell =1}^N \frac{s_i}{d_{i, \ell}} \Re\left( \widetilde{r}_{i, \ell}{\widetilde{\mu}_{\ell}}^* \right) < 0 \right ),
\label{eq25}
\eeqarr
where $\Re\left(\cdot\right)$ indicates the real part operator. Using the statistics derived in (\ref{eq8}) for $\widetilde{\undb{r}}$ conditioned on $s$, $\Re\left( \widetilde{r}_{i, \ell}{\widetilde{\mu}_{\ell}}^* \right)$ follows a normal distribution, i.e., $\Re\left( \widetilde{r}_{i, \ell}{\widetilde{\mu}_{\ell}}^* \right) \!\sim\! {\mathcal{N}}\left( s_i\left| \widetilde{\mu}_{\ell} \right|^2, \frac{\left| \widetilde{\mu}_{\ell} \right|^2d_{i, \ell}}{2}\right)$, which implies that the term $\sum_{\ell =1}^N \frac{s_i}{d_{i, \ell}} \Re\left( \widetilde{r}_{i, \ell}{\widetilde{\mu}_{\ell}}^* \right)$ also follows a normal distribution with mean $\sum_{\ell = 1}^N \frac{s_i^2{\left| \widetilde{\mu}_{\ell} \right|^2}}{d_{i, \ell}}$ and variance $\sum_{\ell = 1}^N \frac{s_i^2{\left| \widetilde{\mu}_{\ell} \right|^2}}{2d_{i, \ell}}$. Thus, for the case of $d_{i, \ell} = d_{j, \ell}$ for $s_i =- s_j$, the following analytical expression for the PEP is deduced:
\beq
P_{i \rightarrow j} \!=\! Q\left(\! \sqrt{\sum_{\ell = 1}^N \frac{2s_i^2{\left| \widetilde{\mu}_{\ell} \right|^2}}{d_{i, \ell}}}\right)\!=\! Q\left(\! \sqrt{\sum\limits_{p=1}^L \frac{2\Gamma_i\lambda_p \sum\limits_{q=1}^{q_p}k_{q,p}}{\Gamma_i\lambda_p + 1}} \right).
\label{eq26}
\eeq
Putting all the above results together, the $P_{i \rightarrow j}$ for all aforementioned three cases is given as follows:
\beqarr
P_{i \rightarrow j}\!\! =\!\! \left\{
\begin{array}{@{}ll@{}}
        F_{\chi_{ij}^2}\left( \alpha_{ij} \right)\!,&  d_{i, \ell} > d_{j, \ell}\\
        1-F_{\chi_{ij}^2}\left( \alpha_{ij} \right)\!,&  d_{i, \ell} < d_{j, \ell}\\
        \!\!Q\!\!\left(\!\! \sqrt{\sum\limits_{p=1}^L\!\!\! \frac{2\Gamma_i\lambda_p \sum\limits_{q=1}^{q_p}k_{q,p}}{\Gamma_i\lambda_p + 1}} \right)\!,&d_{i, \ell} \!= \!d_{j, \ell}\, \text{for}\, s_i \!=\! -s_j\\
  \end{array}.\right.\,\,
\label{eq27}
\eeqarr
The latter expression can be straightforwardly used to obtain the following analytical expression for the union bound on the SEP performance:
\beqarr
P_{e, UB} &\leq & \frac{1}{M} \sum_{i=1}^M 
\left [ \sum_{\substack{j=1, \\ d_{i,\ell} > d_{j,\ell}}}^M\!\!\!\! F_{\chi_{ij}^2}\left( \alpha_{ij} \right)
+ \!\!\!\!\!\sum_{\substack{j=1, \\ d_{i,\ell} < d_{j,\ell}}}^M \!\!\!\left( 1 - F_{\chi_{ij}^2} \left( \alpha_{ij} \right)\right) \right.\nn \\
&&\left.+ \sum_{\substack{j=1, j\neq i, \\ d_{i,\ell} = d_{j,\ell}}}^MQ\left( \sqrt{\sum\limits_{p=1}^L \frac{2\Gamma_i\lambda_p \sum\limits_{q=1}^{q_p}k_{q,p}}{\Gamma_i\lambda_p + 1}} \right)\right ].
\label{eq28}
\eeqarr
\subsection{Tight Approximation for Massive SIMO Systems}
In a massive SIMO system setting, the receiver will be equipped with a large number of antennas, i.e.,  $N \gg 1$. To evaluate the SEP performance for such cases, we hereinafter derive an analytical approximation based on the previously presented union bound.
 The union bound on the SEP, utilizing the PEP for large $N$ denoted by $P_{i \rightarrow j \big |_{N \gg 1}}$, is expressed as:
\beq
P_{e, UB \big |_{N \gg 1}} \leq \frac{1}{M} \sum_{i=1}^M \sum_{\substack{j=1, \\ j\neq i}}^M P_{i \rightarrow j \big |_{N \gg 1}}. 
\label{eq29}
\eeq
To proceed, we consider, as before, the three cases: \textit{i}) $d_{i, \ell} > d_{j, \ell}$; \textit{ii}) $d_{i, \ell} < d_{j, \ell}$; and \textit{iii}) $d_{i, \ell} = d_{j, \ell}$ for $s_i =- s_j$. To this end, we first approximate the distribution of $\chi_{ij}^2$ given in (\ref{eq17}) for the case of large $N$. By applying the central limit theorem, the distribution of $\chi_{ij}^2$ can be approximated as a Gaussian distribution with mean $\mu_{ij,X} \triangleq \sum\limits_{p=1}^L \beta_{ijp}\left( 1 + \gamma_{ijp}\sum\limits_{q=1}^{q_p} k_{q,p}\right)$ and variance $\sigma_{ij,X}^2 \triangleq \sum\limits_{p=1}^L \beta_{ijp}^2\left( 1 + 2\gamma_{ijp}\sum\limits_{q=1}^{q_p} k_{q,p}\right)$, i.e., $
\chi_{ij}^2$ approximately follows $X_{ij} \sim {\mathcal{N}}\left( \mu_{ij,X},  \sigma_{ij,X}^2\right)$. Accordingly, \eqref{eq29} can be re-expressed as follows:
\beqarr
P_{i \rightarrow j \big |_{N \gg 1}}\!\! =\!\! \left\{
\begin{array}{@{}ll@{}}
        \Pr \left( X_{ij} < \alpha_{ij} \right),&  d_{i, \ell} > d_{j, \ell}\\
        \Pr \left( X_{ij} > \alpha_{ij} \right)\!,&  d_{i, \ell} < d_{j, \ell}\\        
  \end{array}\right..
\label{eq30}
\eeqarr
Using the statistical properties of $X_{ij}$ and (\ref{eq26}), the PEP for large $N$ for the aforedescribed three can be obtained as
\beqarr
P_{i \rightarrow j \big |_{N \gg 1}}\!\!\!\! =\!\! \left\{
\begin{array}{@{}ll@{}}
        1-Q\left( \frac{\alpha_{ij} - \mu_{ij,X}}{\sigma_{ij,X}} \right)\!,&  d_{i, \ell} > d_{j, \ell}\\
         Q\left( \frac{\alpha_{ij} - \mu_{ij,X}}{\sigma_{ij,X}} \right)\!,&  d_{i, \ell} < d_{j, \ell}\\
        \!\!Q\!\!\left(\!\! \sqrt{\sum\limits_{p=1}^L\!\!\! \frac{2\Gamma_i\lambda_p \sum\limits_{q=1}^{q_p}k_{q,p}}{\Gamma_i\lambda_p + 1}} \right)\!\!,\!&d_{i, \ell} \!= \!d_{j, \ell}\, \text{for}\, s_i \!=\! -s_j\\
  \end{array}\right.\!\!\!\!\!\!\!\!\!\!.
\eeqarr
By substituting the latter expression into \eqref{eq29}, the following closed-form expression for the union bound on the SEP, for the case of massive SIMO systems, is deduced:
\beqarr
\!\!\!\!\!\!\!\!P_{e, UB \big |_{N \gg 1}} &\leq & \frac{1}{M} \sum_{i=1}^M 
\left [ \sum_{\substack{j=1, \\ d_{i,\ell} > d_{j,\ell}}}^M\!\!\!\! \left( 1-Q\left( \frac{\alpha_{ij} - \mu_{ij,X}}{\sigma_{ij,X}} \right)\right) \right. \nn \\
 && +\left.\!\!\!\!\!\sum_{\substack{j=1, \\ d_{i,\ell} < d_{j,\ell}}}^M \!\!\! Q\left( \frac{\alpha_{ij} - \mu_{ij,X}}{\sigma_{ij,X}} \right) \right.\nn \\
&&\left.+ \sum_{\substack{j=1, j\neq i, \\ d_{i,\ell} = d_{j,\ell}}}^MQ\left( \sqrt{\sum\limits_{p=1}^L \frac{2\Gamma_i\lambda_p \sum\limits_{q=1}^{q_p}k_{q,p}}{\Gamma_i\lambda_p + 1}} \right)\right ].
\eeqarr
%%%%%%%%%%%%%%%%%%%%%%%%%%%%%%%%%%%%%%%%%%%%%%%%%%%%%%%%%%%%%%%%%%%%%%%%%%%%%%%%%%%%%%%%%%
%%%%%%%%%%%%%%%%%%%%%%%%%%%%%%%%%%%%%%%%%%%%%%%%%%%%%%%%%%%%%%%%%%%%%%%%%%%%%%%%%%%%%%%%%%
\section{$M$-Level ASK Modulation Optimization}
This section presents a framework for determining the optimal symbols for both the considered one-sided and two-sided ASK modulation schemes. Our optimization design aims to minimize the union bound on the SEP expression in~\eqref{eq28} under an average transmit energy constraint. To accomplish this, the problem is formulated in terms of the average SNRs, $\Gamma_m$ $\forall m = 1, \ldots, \widetilde{M}$ with $\widetilde{M}$ being equal to $M$ for the one-sided ASK modulation scheme and $M/2$ for the two-sided one, and is mathematically expressed as follows:
\begin{equation*}%\label{eq30}
{\mathcal{OP}}:\,\min_{\Gamma_1,\ldots,\Gamma_{\widetilde{M}}} P_{e,UB} \quad\,\,
\text{s.t. } \sum\limits_{m=1}^{\widetilde{M}} \Gamma_m = {\widetilde{M}} \Gamma_{av}.
\end{equation*}
To solve this constrained optimization problem, we deploy the Lagrangian multiplier technique. The corresponding Lagrangian function is formulated as follows:
\beq
{\mathcal{L}}\left(\Gamma_1, \ldots, \Gamma_{\widetilde{M}}, \eta \right) = P_{e,UB} + \eta\left(\sum\limits_{m=1}^{\widetilde{M}} \Gamma_m - {\widetilde{M}} \Gamma_{av} \right),
\label{eq33}
\eeq
where $\eta$ is the Lagrangian multiplier. To determine the optimal SNR values, denoted as $ \Gamma_{opt, m}$ $\forall m = 1, \ldots, \widetilde{M}$, the following set of differential equations need to be solved simultaneously:
\beqarr
\frac{\partial {\mathcal{L}} \left(\Gamma_1, \ldots, \Gamma_{\widetilde{M}}, \eta \right)}{\partial \Gamma_t} &=& 0 \,\,\,\,  \forall t = 1,\ldots, \widetilde{M} \, , \nn \\
\frac{\partial {\mathcal{L}} \left(\Gamma_1, \ldots, \Gamma_{\widetilde{M}}, \eta \right)}{\partial \eta} &=& 0 \, .
\label{eq34}
\eeqarr
Using the union bound on the SEP given in (\ref{eq12}) in the Lagrangian function defined in (\ref{eq33}), the above differential equations can be reformulated as follows:
\beqarr
\frac{1}{\widetilde{M}} \sum_{i=1}^{\widetilde{M}} \sum_{\substack{j=1, \\ j\neq i}}^{\widetilde{M}} \frac{\partial P_{i \rightarrow j}}{\partial\Gamma_t} + \eta &=&0\,\,\,\, \forall t = 1,\ldots, \widetilde{M} \, , \nn \\
\sum\limits_{m=1}^{\widetilde{M}} \Gamma_m &=& {\widetilde{M}} \Gamma_{av}.
\label{eq35}
\eeqarr
The differentiation of the PEP expression $\frac{\partial P_{i \rightarrow j}}{\partial\Gamma_t}$ in (\ref{eq35}) can be carried out using (\ref{eq27}) as follows:
\beqarr
\frac{\partial P_{i \rightarrow j}}{\partial\Gamma_t}\!\! =\!\! \left\{
\begin{array}{@{}ll@{}}
        \frac{\partial F_{\chi_{ij}^2}\left( \alpha_{ij} \right)}{\partial\Gamma_t}\!,&  d_{i, \ell} > d_{j, \ell}\\
        -\frac{\partial F_{\chi_{ij}^2}\left( \alpha_{ij} \right)}{\partial\Gamma_t}\!,&  d_{i, \ell} < d_{j, \ell}\\
        -\frac{e^{-\frac{\widetilde{\Gamma_i}^2}{2}}}{\sqrt{2\pi\widetilde{\Gamma_i}}}\sum\limits_{p=1}^L \frac{\lambda_p\sum\limits_{q=1}^{q_p}k_{q,p}}{\left( \Gamma_i\lambda_p+1\right)^2}\!,&\,\,\,\,\,t=i\\
  \end{array},\right. 
  \label{eq36}
\eeqarr 
where $\widetilde{\Gamma_i} \triangleq \sum\limits_{p=1}^L \frac{2\Gamma_i\lambda_p\sum\limits_{q=1}^{q_p}k_{q,p}}{\Gamma_i\lambda_p+1}$. Utilizing the m.g.f. expression in (\ref{eq21}), $\frac{\partial F_{\chi_{ij}^2}\left( \alpha_{ij} \right)}{\partial\Gamma_t}$ can be computed as: 
\beqarr
\frac{\partial F_{\chi_{ij}^2}\left( \alpha_{ij} \right)}{\partial\Gamma_t} \!=\! \sum_{u = 0}^{\xi-1} \frac{\left(\xi-1 \right)^u}{x^u u!} \frac{\partial}{\partial \Gamma_t}G_{ij} \left( \nu\right)R_u\left( \nu\right)\Bigg |_{\nu = \frac{1-\xi}{\alpha_{ij}}}\!.
\label{eq37}
\eeqarr
To derive $\frac{\partial}{\partial \Gamma_t}G_{ij} \left( \nu\right)R_u\left( \nu\right)$, we first use the property:
\beqarr
\!\!\!\!\!\!\frac{\partial G_{ij} \left( \nu\right)R_u\left( \nu\right)}{\partial \Gamma_t} = G_{ij} \left( \nu\right)\frac{\partial R_u\left( \nu\right)}{\partial \Gamma_t} + R_u\left( \nu\right) \frac{\partial G_{ij} \left( \nu\right)}{\partial \Gamma_t}.
\label{eq38}
\eeqarr
To compute the derivative $\frac{\partial R_u\left( \nu\right)}{\partial \Gamma_t}$, we use the expressions provided in (\ref{eq23}), yielding:
\beqarr
\!\!\!\!\!\!\!\!\!\!\!\!\!\!\!\!\!\!\!\!\!\!\!\!\!\!\!\!\!\!\!\!\!\!\!\!\frac{\partial R_u\left( \nu\right)}{\partial \Gamma_t} &=& \sum_{v=0}^{u-1}\binom{u-1}{v}\left(g^{(u-1-v)}\left( \nu\right)\frac{\partial R_v\left( \nu\right)}{\partial \Gamma_t} \right.\nn\\
&&\left. +  R_v\left(\nu \right)\frac{\partial g^{(u-1-v)}\left( \nu\right)}{\partial \Gamma_t}\right), 
\label{eq39}
\eeqarr
It can be observed from (\ref{eq23}) that $g^{(n)}\left( \nu \right)$ is a function of $\beta_{ij,p}, \gamma_{ij,p}$, and $\nu$, all of which depend on $\Gamma_t$. Thus, by applying the chain rule, the derivative of 
$g^{(n)}\left( \nu \right)$ with respect to $\Gamma_t$ can be expressed as follows:
\beqarr
\!\!\!\!\!\!\!\!\!\!\!\!\!\!\!\!\!\!\!\!\!\!\!\!\!\!\!\!\!\!\!\!\!\!\!\!\frac{\partial g^{(n)}\left( \nu\right)}{\partial \Gamma_t} &=& \frac{\partial g^{(n)}\left( \nu\right)}{\partial \beta_{ij,p}}\frac{\partial \beta_{ij,p}}{\partial \Gamma_t} + \frac{\partial g^{(n)}\left( \nu\right)}{\partial \gamma_{ij,p}}\frac{\partial \gamma_{ij,p}}{\partial \Gamma_t} \nn \\
&&+ \frac{\partial g^{(n)}\left( \nu\right)}{\partial \nu}\frac{\partial \nu}{\partial \Gamma_t},
\label{eq40}
\eeqarr
where we can easily obtain the following expressions:
\beqarr
&&\!\!\!\!\!\!
\frac{\partial g^{(n)}\left( \nu\right)}{\partial \beta_{ij,p}} = \sum \limits_{p=1}^L \left [ \frac{n!\left(n+1\right)\beta_{ij,p}^{n}}{\left(1-\beta_{ij,p}\nu \right)^{n+3}} \right. \nn \\
&& \! \! \! \!
\left.\times \left(\gamma_{ij,p}\left(n+1+\beta_{ij,p}\nu\right)\sum_{q=1}^{q_p}k_{q,p} + q_p\left(1-\beta_{ij,p}\nu\right) \right)\right ], \nn \\
&&\!\!\!\!\!\!
\frac{\partial g^{(n)}\left( \nu\right)}{\partial \gamma_{ij,p}} = \sum\limits_{p=1}^L\frac{n!\left(n+1\right)\sum_{q=1}^{q_p}k_{q,p}\beta_{ij,p}^{n+1}}{\left(1-\beta_{ij,p}\nu\right)^{n+2}}, \nn \\
&&\!\!\!\!\!\!
\frac{\partial g^{(n)}\left( \nu\right)}{\partial \nu} = \sum\limits_{p=1}^L \frac{n!\left(n+1\right)\beta_{ij,p}^{n+2}}{\left( 1 - \beta_{ij,p}\nu\right)^{n+3}}\nn \\
&& \qquad \quad \times\left [ \left(n+2\right)\gamma_{ij,p}\sum_{q=1}^{q_p}k_{q,p} + q_p\left(1-\beta_{ij,p}\nu\right)\right].
\label{eq41}
\eeqarr

To now compute $\frac{\partial G_{ij} \left( \nu\right)}{\partial \Gamma_t}$ in~\eqref{eq38}, we first consider the function
\beq
f(\beta_{ij,p}, \gamma_{ij,p}, \nu) \triangleq  \frac{\exp \left \{ \frac{\nu \beta_{ij,p} \gamma_{ij, p} \sum \limits_{q=1}^{q_p}k_{q,p}}{1-\nu\beta_{ij,p}}\right\}}{\left(1-\nu\beta_{ij,p} \right)^{q_p}},
\label{eq42}
\eeq
and utilize the expression for $G_{ij}(\nu)$ given in (\ref{eq19}). Then, the derivative of $G_{ij} \left( \nu\right)$ with respect to $\Gamma_t$ is given by:
\beqarr
\!\!\frac{\partial G_{ij} \left( \nu\right)}{\partial \Gamma_t}\!\! = \!\!\sum_{p=1}^L \left [ \frac{\partial f(\beta_{ij,p}, \gamma_{ij,p}, \nu)}{\partial \Gamma_t} \prod\limits_{\substack{c=1 \\ c \neq p}}^L f(\beta_{ij,c}, \gamma_{ij,c}, \nu) \right]\!\!.
\label{eq43}
\eeqarr
Using the chain rule, yields:
\beqarr
&&\!\!\!\!\!\!\!\!\!\!\!\!\!\!\frac{\partial f(\beta_{ij,p}, \gamma_{ij,p}, \nu) }{\partial \Gamma_t} = \frac{\partial f(\beta_{ij,p}, \gamma_{ij,p}, \nu)}{\partial \beta_{ij,p}}\frac{\partial \beta_{ij,p}}{\partial \Gamma_t}\nn  \\
&& \!\!\!+ \frac{\partial f(\beta_{ij,p}, \gamma_{ij,p}, \nu)}{\partial \gamma_{ij,p}}\frac{\partial \gamma_{ij,p}}{\partial \Gamma_t} + \frac{\partial f(\beta_{ij,p}, \gamma_{ij,p}, \nu)}{\partial \nu}\frac{\partial \nu}{\partial \Gamma_t},
\label{eq44}
\eeqarr
where the involved partial derivatives are given by:
\beqarr
&&\!\!\!\!\!\!\!\frac{\partial f(\beta_{ij,p}, \gamma_{ij,p}, \nu)}{\partial \beta_{ij,p}} = \frac{e^{ \frac{\nu \beta_{ij,p} \gamma_{ij, p} \sum\limits_{q=1}^{q_p}k_{q,p}}{1-\nu\beta_{ij,p}}}\left(1-\nu\beta_{ij,p} \right)^{-q_p-2}}{\gamma_{ij,p}\sum\limits_{q=1}^{q_p}k_{q,p}+q_p\left(1-\nu\beta_{ij,p}\right)},\nn \\
&&\!\!\!\!\!\!\!\frac{\partial f(\beta_{ij,p}, \gamma_{ij,p}, \nu)}{\partial \gamma_{ij,p}} = \frac{e^ { \frac{\nu \beta_{ij,p} \gamma_{ij, p} \sum\limits_{q=1}^{q_p}k_{q,p}}{1-\nu\beta_{ij,p}}}\sum\limits_{q=1}^{q_p}k_{q,p}}{\left(\nu \beta_{ij,p}   \right)^{-1}\left(1-\nu\beta_{ij,p} \right)^{q_p + 1}}, \nn \\
&&\!\!\!\!\!\!\!\frac{\partial f(\beta_{ij,p}, \gamma_{ij,p}, \nu)}{\partial \nu} = \frac{\beta_{ij,p}e^{ \frac{\nu \beta_{ij,p} \gamma_{ij, p} \sum\limits_{q=1}^{q_p}k_{q,p}}{1-\nu\beta_{ij,p}}}}{\left(1-\nu\beta_{ij,p} \right)^{q_p + 2}} \nn \\
&&\qquad\qquad\qquad\quad\times\!\! \left(\gamma_{ij,p}\sum_{q=1}^{q_p}k_{q,p} + q_p \left( 1 - \beta_{ij,p}\nu\right) \right)\!\!.
\label{eq45}
\eeqarr
The expressions for the terms $\frac{\partial \beta_{ij,p}}{\partial \Gamma_t}, \frac{\partial \gamma_{ij,p}}{\partial \Gamma_t}, $ and $\frac{\partial \alpha_{ij}}{\partial \Gamma_t}$ for $t = i$ are given as follows:
\beqarr
&&\!\!\!\!\!\!\!\!\!\!\!\!\!\!\frac{\partial \beta_{ij,p}}{\partial \Gamma_t} = \frac{\lambda_p}{\Gamma_j \lambda_p +1}, \nn \\
&&\!\!\!\!\!\!\!\!\!\!\!\!\!\!\frac{\partial \gamma_{ij,p}}{\partial \Gamma_t} = \frac{\sqrt{\Gamma_i\lambda_p}\phi_i - \sqrt{\Gamma_j\lambda_p}\phi_j}{\left( \Gamma_i - \Gamma_j\right)^3\lambda_p^2\sqrt{\Gamma_i\lambda_p}}\left( 2\sqrt{\Gamma_i\lambda_p}\sqrt{\Gamma_j\lambda_p}\phi_j \right. \nn\\
&&\quad\left. + \left( \Gamma_i\lambda_p\right)^{3/2} \sqrt{\Gamma_i\lambda_p}\phi_j +\sqrt{\Gamma_i\lambda_p}\left( \Gamma_j\lambda_p\right)^{3/2}\phi_j\right.\nn \\
&&\quad\left.- \lambda_p\left( \Gamma_i+\Gamma_j +2\Gamma_i\Gamma_j\lambda_p\right)\phi_i\right), \nn \\
&&\!\!\!\!\!\!\!\!\!\!\!\!\!\!\frac{\partial \alpha_{ij}}{\partial \Gamma_t} = \sum_{p =1}^L \left( \left( \frac{\lambda_pq_p}{\Gamma_i\lambda_p + 1} \prod\limits_{\substack{c=1\\c\neq p}}^L \ln \left( \Gamma_i\lambda_c+1\right)^{q_c}\right)\right. \nn \\
&&+\!\!\!\left.\frac{\left(\sqrt{\Gamma_i\Gamma_j}\lambda_p\phi_j -\Gamma_j\lambda_p\phi_i\right)\left( \Gamma_i -\Gamma_j\right)^{-2} \sum\limits_{q=1}^{q_p}k_{q,p}}{\Gamma_i^{-1} \left(\Gamma_i\lambda_p\right)^{3/2}\left(\sqrt{\Gamma_i\lambda_p}\phi_i - \sqrt{\Gamma_j\lambda_p}\phi_j \right)^{-1}}\!\!\!\right)\!\!,
\label{eq46}
\eeqarr
and, for $t = j$, as:
\beqarr
&&\!\!\!\!\!\!\!\!\!\!\!\!\frac{\partial \beta_{ij,p}}{\partial \Gamma_t} = -\frac{\lambda_p\left( 1+ \Gamma_i \lambda_p\right)}{\left(\Gamma_j \lambda_p +1\right)^2}, \nn \\
&&\!\!\!\!\!\!\!\!\!\!\!\!\frac{\partial \gamma_{ij,p}}{\partial \Gamma_t} = 
\frac{\left(\Gamma_i\lambda_p +1 \right)\left( \sqrt{\Gamma_i\lambda_p}\phi_i - \sqrt{\Gamma_j\lambda_p}\phi_j\right)}{\left( \Gamma_j - \Gamma_i\right)^3\lambda_p^2\sqrt{\Gamma_j\lambda_p}}\nn \\
&& \quad \,\,\,\,\times \left( \left( \Gamma_i + \Gamma_j\right)\lambda_p \phi_j - 2\sqrt{\Gamma_i\lambda_p}\sqrt{\Gamma_j\lambda_p}\phi_i\right), \nn\\
&&\!\!\!\!\!\!\!\!\!\!\!\!\frac{\partial \alpha_{ij}}{\partial \Gamma_t} = \sum\limits_{p =1}^L \left( -\left( \frac{\lambda_pq_p}{\Gamma_j\lambda_p + 1} \prod\limits_{\substack{c=1\\c\neq p}}^L \ln \left( \Gamma_j\lambda_c+1\right)^{q_c}\right)\right. \nn \\
&&\!+\!\!\left.\frac{\left(\sqrt{\Gamma_i\Gamma_j}\lambda_p\phi_i -\Gamma_j\lambda_p\phi_j\right)\left( \Gamma_i -\Gamma_j\right)^{-2} \sum\limits_{q=1}^{q_p}k_{q,p}}{\Gamma_j^{-1} \left(\Gamma_j\lambda_p\right)^{3/2}\left(\sqrt{\Gamma_i\lambda_p}\phi_i - \sqrt{\Gamma_j\lambda_p}\phi_j \right)^{-1}}\!\!\right)\!\!.
\label{eq47}
\eeqarr

Finally, solving the set of differential equations in (\ref{eq34}) using the derivations in (\ref{eq36})--(\ref{eq47}), yields the optimal SNR values for $\mathcal{OP}$. Due to the high nonlinearity of the equations, the system of differential equations is solved numerically.
%%%%%%%%%%%%%%%%%%%%%%%%%%%%%%%%%%%%%%%%%%%%%%%%%%%%%%%%%%%%%%%%%%%%%%%%%%%%%%%%%%%%%%%%%%

\section{Numerical Results and Discussions}
In this section, we numerically evaluate the error performance of conventional one- and two-sided ASK modulation schemes, as well as their optimal counterparts derived using the proposed optimization framework presented in Section~IV, for the considered noncoherent SIMO systems over i.i.d. channels, exponentially correlated, and uniformly correlated Rician fading channels.

\begin{figure}[h]
\centering
    \includegraphics[height=2.5in,width=3.5in]{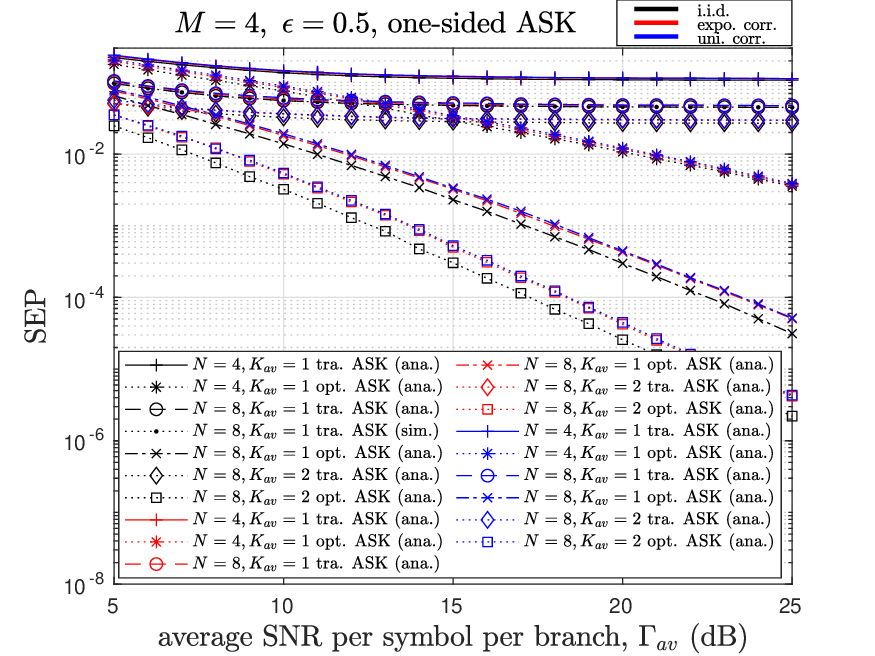}
    \caption{SEP performance versus the average SNR per symbol per branch for $4$-level one-sided ASK modulation over i.i.d., exponentially correlated, and uniformly correlated fading channels, considering $K_{av} =\{1, 2\}$, $\epsilon = 0.5$, and $N =\{4, 8\}$.}
    \label{f_1}
\end{figure}
\begin{figure}[h]
\centering
    \includegraphics[height=2.5in,width=3.5in]{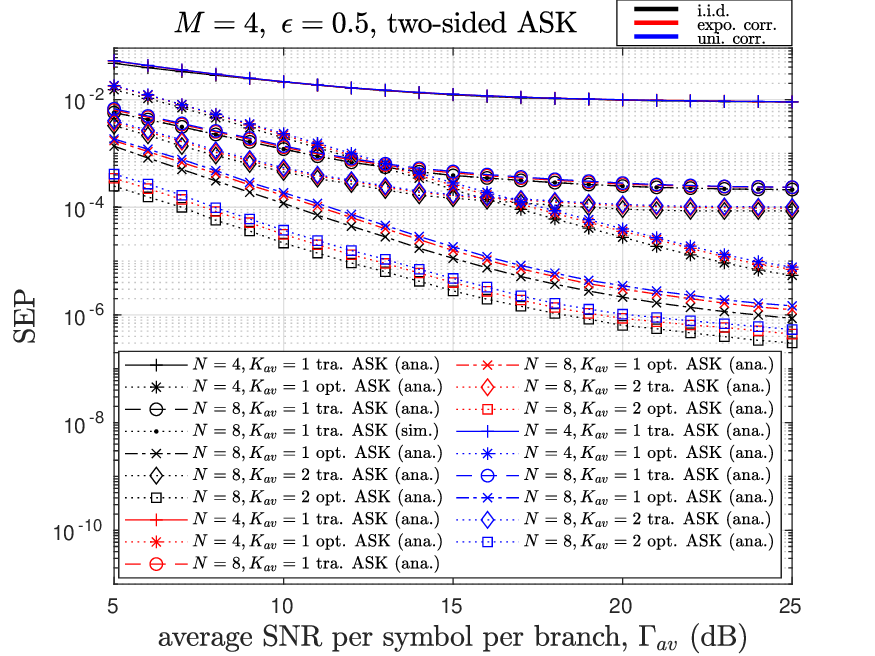}
    \caption{Similar to Fig.~\ref{f_1} but for $4$-level two-sided ASK modulation.}
    \label{f_2}
\end{figure}
\begin{figure}[h]
\centering
    \includegraphics[height=2.5in,width=3.5in]{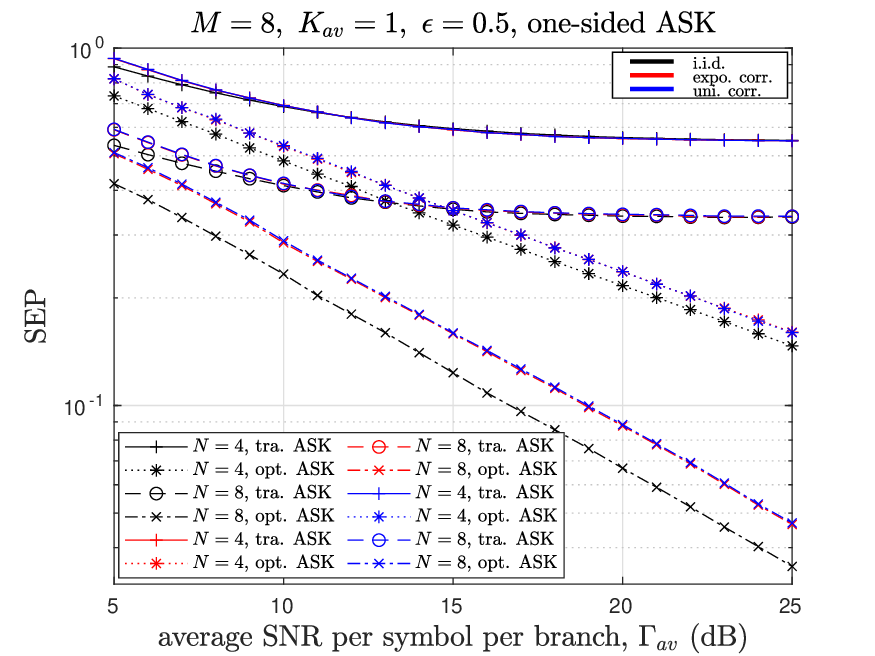}
    \caption{SEP performance versus the average SNR per symbol per branch for $8$-level one-sided ASK modulation over i.i.d., exponentially correlated, and uniformly correlated channels, considering $K_{av} = 1$, $\epsilon = 0.5$, and $N =\{4, 8\}$.}
    \label{f_3}
\end{figure}
\begin{figure}[h]
\centering
    \includegraphics[height=2.5in,width=3.5in]{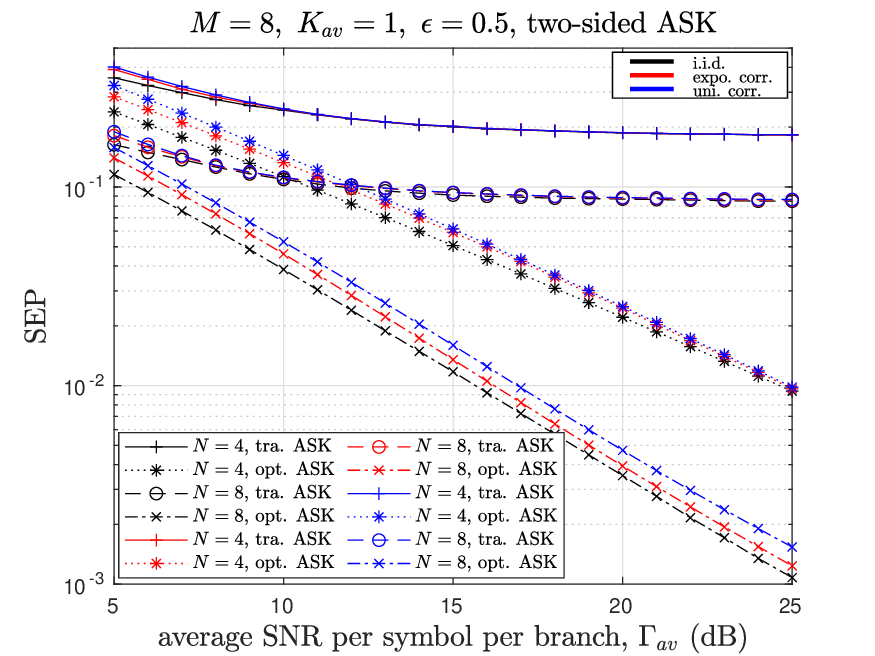}
    \caption{Similar to Fig.~\ref{f_3} but for $8$-level two-sided ASK modulation.}
    \label{f_4}
\end{figure}
Figs.~\ref{f_1} and~\ref{f_2} present SEP results for $4$-level one- and two-sided ASK modulations, respectively, considering the correlation factor $\epsilon = 0.5$. The performance was evaluated for varying numbers of receive antennas $N=\{4,8\}$ and average Rician factors $K_{av}=\{1,2\}$. For the analysis, the SEP plots were obtained using the union bound expression given in (\ref{eq28}), ensuring that the truncation error in the computed SEP remains below $0.1\%$ by selecting a sufficiently large value for $\xi$. It can be observed that, for the system employing traditional equispaced one- and two-sided ASK modulation schemes, the SEP values saturate at their respective minimum values as the average SNR increases. However, when optimal ASK modulation schemes are used, the system achieves better error performance. This performance improvement is shown to increase as the number of receive antennas $N$ increases. This trend, however, does not occur with the two-sided ASK modulation scheme. Furthermore, the system using traditional equispaced two-sided ASK modulation schemes performs better than the system using traditional equispaced one-sided ASK modulation schemes. 

Figs.~\ref{f_3} and \ref{f_4} present similar SEP results with Figs.~\ref{f_1} and \ref{f_2} for one- and two-sided ASK modulation schemes with a modulation order of $M=8$. It is demonstrated that the error performance degrades as $M$ increases, owing to the fact that higher-order modulation schemes exhibit lower error performance. However, the overall error performance trends remain consistent with those observed for $M=4$. Moreover, the system employing both one- and two-sided modulation schemes exhibits superior error performance in i.i.d. channels compared to exponentially correlated channels, which, in turn, perform better than uniformly correlated channels. This trend is consistently observed across the modulation orders $M=4$ and $8$, and holds for both traditional and optimal ASK modulation schemes. Furthermore, owing to the fact that the SEP values decrease as $N$ and the average Rician factor increase for any given average SNR value. 

\begin{figure}[h]
\centering
    \includegraphics[height=2.5in,width=3.5in]{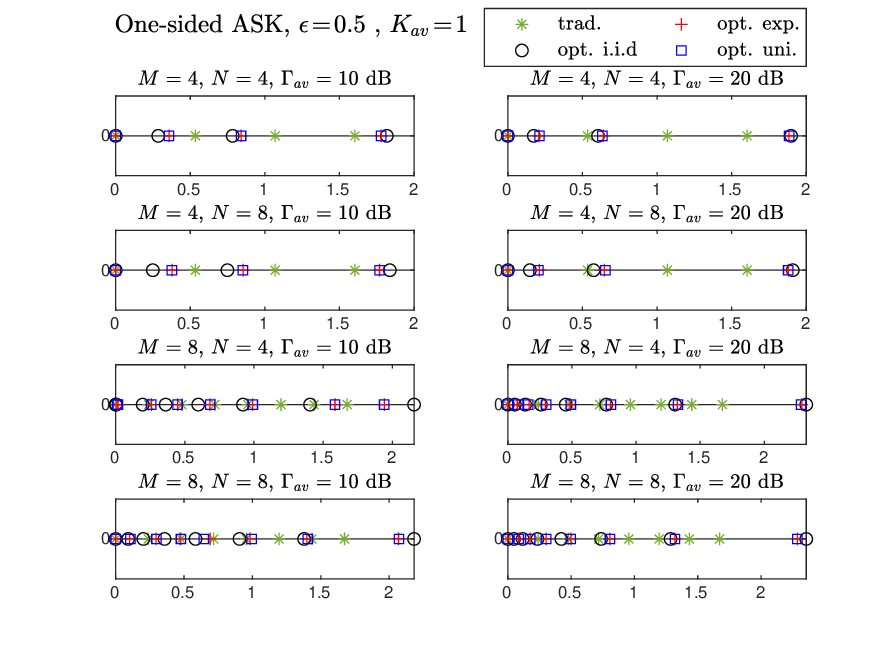}
    \caption{Traditional and the proposed optimal $M$-level one-sided ASK constellation diagrams with $M =\{4, 8\}$, $\Gamma_{av} =\{10, 20\}$ dB, and $N =\{4, 8\}$.}
    \label{f_5}
\end{figure}
\begin{figure}[h]
\centering
    \includegraphics[height=2.5in,width=3.5in]{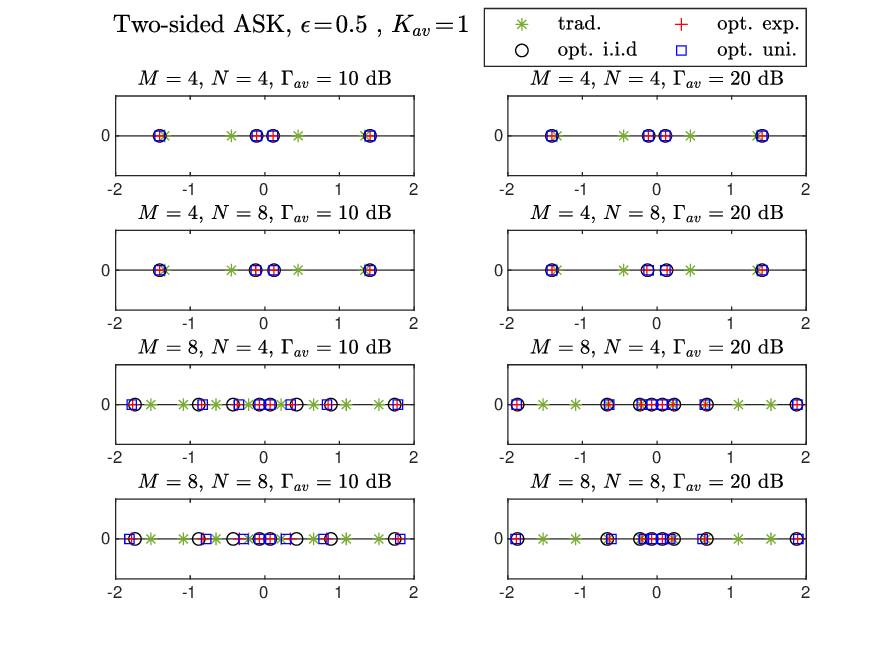}
    \caption{Similar to Fig.~\ref{f_5} but for $M$-level two-sided ASK modulation.}
    \label{f_6}
\end{figure}

The constellation diagrams for one- and two-sided ASK modulation schemes are presented respectively in Figs.~\ref{f_5} and~\ref{f_6}, comparing the traditional equispaced ASK modulation with our proposed optimal ASK modulation schemes. For a fair comparison, the constellation points have been normalized with the average system energy $E_{av}$. It is shown that the optimal constellation points of one- and two-sided ASK modulation schemes differ from the traditional equispaced ASK constellation points. The gap between the optimal points is not uniform and is increasing towards higher-energy constellation points. Moreover, an increase in the number of receive antennas $N$ reduces the gap between the lower-energy constellation points. Higher SNR values further diminish this gap, leading to more compact constellation diagrams. Additionally, as the $M$ increases, the gap between the constellation points decreases further. Furthermore, the type of channel correlation also plays a significant role in shaping the constellation patterns, with the impact being more pronounced in i.i.d. channels compared to exponentially or uniformly correlated channels.
\begin{figure}[h]
\centering
    \includegraphics[height=2.5in,width=3.5in]{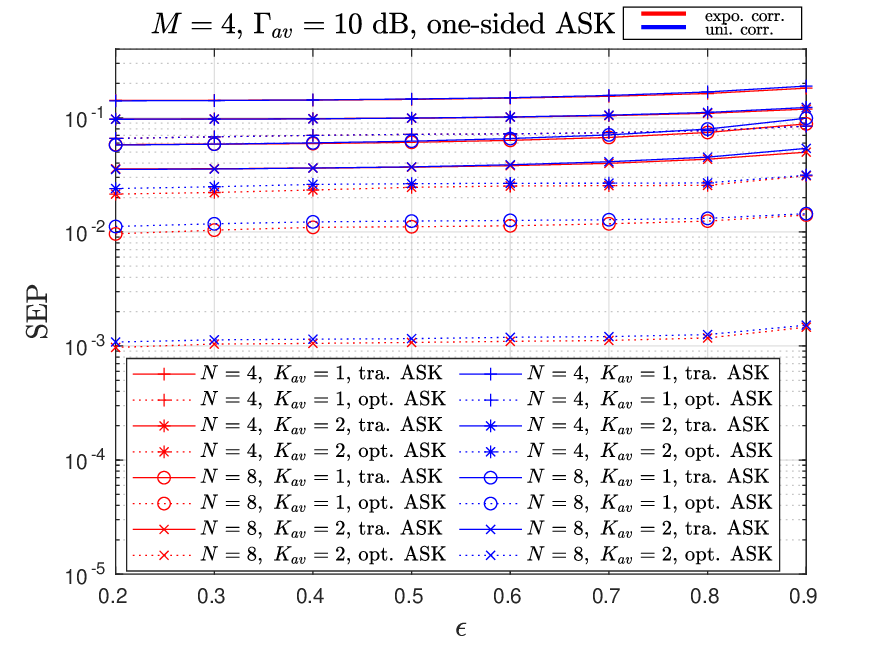}
    \caption{SEP performance versus the channel correlation coefficient for $4$-level one-sided ASK modulation over exponentially correlated and uniformly correlated channels, considering $K_{av} =\{1, 2\}$, $\Gamma_{av} = 10$ dB, and $N =\{4, 8\}$.}
    \label{f_7}
\end{figure}
\begin{figure}[h]
\centering
    \includegraphics[height=2.5in,width=3.5in]{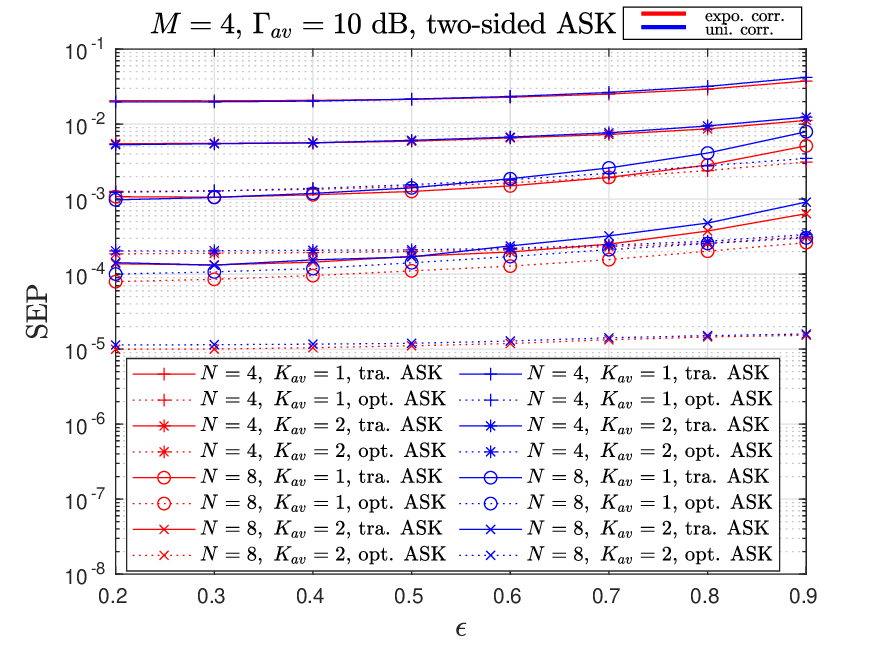}
    \caption{Similar to Fig.~\ref{f_7} but for $4$-level two-sided ASK modulation.}
    \label{f_8}
\end{figure}

Finally, Figs. \ref{f_7} and \ref{f_8} illustrate the variation in SEP performance with respect to the channel correlation coefficient $\epsilon$ for one- and two-sided $4$-level ASK modulation schemes, respectively, considering $N =\{4, 8\}$, $K_{av} =\{1, 2\}$, and $\Gamma_{av} = 10$~dB. The results showcase that, as $\epsilon$ increases, the error performance of the system degrades for both traditional and optimal one- and two-sided ASK modulation schemes. The variation in SEP across different correlation coefficient values is more pronounced using traditional modulation schemes compared to optimal modulation schemes. Moreover, this variation is further increased with an increase in $N$ and the average Rician factor. Consistently with the previous findings, the system employing our proposed optimal modulation schemes achieves lower SEP compared to using traditional equispaced modulation schemes across all $\epsilon$ values. Additionally, increasing the average Rician factor and $N$ further improves error performance by reducing SEP across all $\epsilon$ values.
%%%%%%%%%%%%%%%%%%%%%%%%%%%%%%%%%%%%%%%%%%%%%%%%%%%%%%%%%%%%%%%%%%%%%%%%%%%%%%%%%%%%%%%%%%
%%%%%%%%%%%%%%%%%%%%%%%%%%%%%%%%%%%%%%%%%%%%%%%%%%%%%%%%%%%%%%%%%%%%%%%%%%%%%%%%%%%%%%%%%%
%%%%%%%%%%%%%%%%%%%%%%%%%%%%%%%%%%%%%%%%%%%%%%%%%%%%%%%%%%%%%%%%%%%%%%%%%%%%%%%%%%%%%%%%%%
%%%%%%%%%%%%%%%%%%%%%%%%%%%%%%%%%%%%%%%%%%%%%%%%%%%%%%%%%%%%%%%%%%%%%%%%%%%%%%%%%%%%%%%%%%
\section{Conclusions}
This paper studied noncoherent SIMO wireless communications in which the transmitter employs one- and two-sided ASK modulation schemes for data transmission over correlated Rician fading channels. An optimal noncoherent ML receiver structure was derived that uses the availability of statistical channel knowledge for data detection. Using this receiver structure, a novel expression for the union bound on the SEP performance was obtained. We also devised a framework to determine the optimal one- and two-sided ASK modulation schemes by minimizing the SEP under average transmit energy constraints. It was demonstrated that the system employing traditional equispaced ASK modulation schemes exhibits error performance saturation as the average SNR increases. Moreover, it was shown that the system employing the equispaced two-sided ASK modulation scheme achieves superior error performance compared to its one-sided counterpart. Furthermore, the system utilizing the optimized modulation schemes demonstrated improved error performance relative to traditional equispaced ASK modulation schemes. The error performance was also shown to be superior in i.i.d. channels compared to exponentially correlated channels, which, in turn, outperform uniformly correlated channels. The optimal constellation points for both one- and two-sided ASK modulation schemes were shown to not being uniformly spaced. The gaps between lower-energy constellation points were more sensitive to key system parameters, including the number of receive antennas, the average SNR, and the modulation order. Additionally, it was shown that the type of channel correlation significantly influences the constellation pattern. As the channel correlation coefficient increased, the system's error performance deteriorated. Notably, the error performance of systems employing traditional modulation schemes was more sensitive to changes in the channel correlation compared to those utilizing optimal modulation schemes.

%%%%%%%%%%%%%%%%%%%%%%%%%%%%%%%%%%%%%%%%%%%%%%%%%%%%%%%%%%%%%%%%%%%%%%%%%%%%%%%%%%%%%%%%%%
%%%%%%%%%%%%%%%%%%%%%%%%%%%%%%%%%%%%%%%%%%%%%%%%%%%%%%%%%%%%%%%%%%%%%%%%%%%%%%%%%%%%%%%%%%
%%%%%%%%%%%%%%%%%%%%%%%%%%%%%%%%%%%%%%%%%%%%%%%%%%%%%%%%%%%%%%%%%%%%%%%%%%%%%%%%%%%%%%%%%%
%%%%%%%%%%%%%%%%%%%%%%%%%%%%%%%%%%%%%%%%%%%%%%%%%%%%%%%%%%%%%%%%%%%%%%%%%%%%%%%%%%%%%%%%%%

%%%%%%%%%%%%%%%%%%%%%%%%%%%%%%%%%%%%%%%%%%%%%%%%%%%%%%%%%%%%%%%%%%%%%%%%%%%%%%%%%%%
\end{document}